\documentclass[mathpazo,a4paper]{aamm}
\renewcommand\thisnumber{x}
\renewcommand\thisyear {200x}
\renewcommand\thismonth{xxx}
\renewcommand\thisvolume{xx}

\usepackage{enumerate}

\newcommand{\bit}{\mathrm{bit}}

\newcommand{\rhor}{\rho^{\circ}}
\newcommand{\rhoin}{\rho^{\mathrm{\tiny{in}}}} 
\newcommand{\rhoout}{\rho^{\mathrm{\tiny{out}}}} 

\newcommand{\dd}{d}

\newcommand{\taub}{{\tau_{\mathrm{bulk}}}}

\newcommand{\dt}{\Delta t} 
\newcommand{\dx}{\Delta x}

\newcommand{\uvel}{\vbf{u}}
\newcommand{\ui}{\vv{u}{\xi}}
\newcommand{\uj}{\vv{u}{\xj}}
\newcommand{\uk}{\vv{u}{\xk}}

\newcommand{\uveleq}{\vbf{u}^{*}}

\newcommand{\ci}{\mathbf{c}_{\ii}}

\newcommand{\ii}{\dot{\imath}}
\newcommand{\jj}{\dot{\jmath}}

\newcommand{\x}{\vbf{x}}
\renewcommand{\xi}{x_{\ai}}
\newcommand{\xj}{x_{\aj}}
\newcommand{\xk}{x_{\ak}}

\newcommand{\ek}{\mathbf{e}_{\ak}}

\newcommand{\Lxi}{X_{\ai}}
\newcommand{\Lxj}{X_{\aj}}
\newcommand{\Lxk}{X_{\ak}}

\newcommand{\Lxin}{X_{\mc{I}}}
\newcommand{\Lxout}{X_{\mc{O}}}
\newcommand{\Lxinout}{X_{\mc{I}/\mc{O}}}
\newcommand{\Lxs}{X_{\mc{S}}}

\newcommand{\Cinp}{\mc{C}(\Lxin+1)}

\newcommand{\Coutm}{\mc{C}(\Lxk-\Lxout)}

\newcommand{\bepsilon}{\boldsymbol{\epsilon}}

\newcommand{\ai}{1}
\newcommand{\aj}{2}
\newcommand{\ak}{3}

\newcommand{\fzv}{\vbf{b}}
\newcommand{\fz}{b}

\newcommand{\bfM}{\vbf{M}}

\newcommand{\bfhS}{\vbf{\check{S}}}

\newcommand{\n}{n}
\newcommand{\bfn}{\vbf{\n}}

\renewcommand{\ni}{\n_{\ii}}
\newcommand{\nj}{\n_{\jj}}

\newcommand{\nq}{\n^{\mathrm{\tiny{eq}}}}

\newcommand{\nqi}{\nq_{\ii}}
\newcommand{\nqj}{\nq_{\jj}}

\newcommand{\m}{m}
\newcommand{\bfm}{\vbf{\m}}

\newcommand{\bfmq}{{\bfm}^{\mathrm{\tiny{eq}}}}

\newcommand{\ww}[1]{\omega_{\text{\tiny{#1}}}}
\newcommand{\wci}{\ww{$\ci$}}

\newcommand{\cs}{{c_{\mathrm{s}}}}

\newcommand{\theo}{\text{th}}

\newcommand{\dpl}{(\nabla p)_{\xk}}

\newcommand{\BCC}{\ensuremath{\textsl{BCC}}}

\newcommand{\LB}{\ensuremath{\text{LB}}}
\newcommand{\BGK}{\ensuremath{\text{BGK}}}
\newcommand{\LBBGK}{\ensuremath{\text{LB-BGK}}}
\newcommand{\MRT}{\ensuremath{\text{MRT}}}
\newcommand{\LBMRT}{\ensuremath{\text{LB-MRT}}}

\newcommand{\BC}{\ensuremath{\text{BC}}}
\newcommand{\ICh}{\ensuremath{\text{I-Ch}}}
\newcommand{\pBC}{\ensuremath{\text{$p$-BC}}}
\newcommand{\dpls}{\ensuremath{\text{$\nabla p$-S}}}

\newcommand{\etal}{\ensuremath{\textsl{et al.}}}

\newcommand{\mc}[1]{\mathcal{#1}}
\newcommand{\avrg}[1]{\langle{#1}\rangle}

\newcommand{\pwr}[1]{\!\times\!10\sp{#1}}
\newcommand{\pwrr}[1]{10\sp{#1}}
\newcommand{\fracd}[2]{
\displaystyle{\frac{{\displaystyle{#1}}}{{\displaystyle{#2}}}}}

\newcommand{\vbf}[1]{\mathbf{#1}}
\newcommand{\vv}[2]{\mathrm{#1}_{#2}}

\begin{document}


\markboth{Narv\'{a}ez and Harting}{Evaluation of pressure BC for permeability
  calculations}

\title{Evaluation of pressure boundary conditions for
 permeability calculations using the lattice-Boltzmann method}

 \author[Narv\'{a}ez and Harting]{A. Narv\'{a}ez\affil{1}\comma \affil{2}
 and J. Harting\affil{1}\comma \affil{2}\comma\corrauth}
 \address{
\affilnum{1} Department of Applied Physics, Eindhoven University of
Technology,\\ Postbus 513, 5600MB Eindhoven, The Netherlands\\
\affilnum{2} Institute Computational Physics, University of Stuttgart,\\
Pfaffenwaldring 27, 70569 Stuttgart, Germany}

 \email{{\tt j.harting@tue.nl} (J. Harting)}

\begin{abstract}
Lattice-Boltzmann (\LB) simulations are a common tool to numerically
estimate the permeability of porous media. For valuable results, the
porous structure has to be well resolved resulting in a large
computational effort as well as high memory demands. In order to estimate
the permeability of realistic samples, it is of importance to not only
implement very efficient codes, but also to choose the most appropriate
simulation setup to achieve accurate results. With the focus on accuracy
and computational effort, we present a comparison between different
methods to apply an effective pressure gradient, efficient boundary
conditions, as well as two \LB\ implementations based on pore-matrix and
pore-list data structures. 
\end{abstract}



\maketitle

\section{Introduction}
The lattice-Boltzmann (\LB) method is a numerical scheme that is able to
simulate the hydrodynamics of fluids with complex interfacial dynamics and
boundaries~\cite{bib:mcnamara-zanetti,bib:higuera-succi-benzi,
  bib:chen-chen-matthaeus,bib:qian-dhumieres-lallemand,
  bib:chen-chen-martinez-matthaeus,bib:succi-01}. Its popularity stems
from the broad field of possible application and a fair implementation
effort compared to other CFD methods. Unlike schemes that are based on a
discretization of the Navier-Stokes equations, and therefore represent
balance equations at the continuum level (macroscopic), the \LB\ method
represents the dynamics at mesoscopic level by solving the discretized
Boltzmann equation.

There is an increasing interest in the \LB\ method for
simulation of flow in complex geometries since the end of the
1980's~\cite{citeulike:3986220} when hydrodynamic simulation methods were
dominated by finite element schemes that solved the Stokes
equation~\cite{PhysRevB.46.6080,1996AREPS..24...63V}. With the advent of more
powerful computers it became possible to perform detailed simulations of flow
in artificially generated geometries~\cite{koponen:3319}, tomographic
reconstructions of sandstone
samples~\cite{bib:ferreol-rothman,Martys99largescale,MAKHT02,
bib:jens-venturoli-coveney:2004,Ahren06}, or fibrous sheets of
paper~\cite{koponen:716}. An important property estimated using the \LB\
method in those geometries is the permeability~\cite{bib:darcy}, and since the
porous structure has to be well resolved in order to obtain valuable results,
a large computational effort as well as high memory demands is required.
Therefore, it is important to develop very efficient simulation paradigms.

Different alternative simulation setups have been proposed for permeability
estimation which differ in the computational domain setup, how the fluid is
driven, or how an effective pressure gradient is being estimated. Further
possible differences include the choice between single relaxation and
multirelaxation time lattice Boltzmann implementations or data structures
based on a 3D array containing the whole discretized simulation volume
including rock matrix and pore space (pore-matrix) in contrast to data
structures limited to a connected list of pore nodes
(pore-list)~\cite{bib:succi-01,2005ASIM:DXHHW}. 

The current paper focuses on a detailed comparison of some of these possible
implementation details to accurately estimate the permeability of porous media
with the \LB\ method. We compare the well known D3Q19 single relaxation
(\LBBGK)~\cite{bib:bgk,bib:chen-chen-martinez-matthaeus} and multirelaxation
time (\LBMRT)~\cite{2002RSPTA.360..437D} models together with three
alternative setups to estimate the permeability utilizing an injection channel
of variable length (\ICh), pressure boundary conditions (\pBC), or a force
density applied over the sample (\dpls). The geometries being investigated
are a 3D Poiseuille flow in a square pipe and a \BCC\ sphere array. While
the first one has the advantage of a minimal discretization error, the
second one more realistically resembles a natural porous medium.  We also
present a comparison of the efficiency of the \LB-codes based on the above
mentioned pore-matrix and pore-list data structures~\cite{2005ASIM:DXHHW}. 

\section{The Lattice-Boltzmann Method}
\begin{figure}[t]
\begin{center}
\includegraphics[width=0.6\columnwidth]{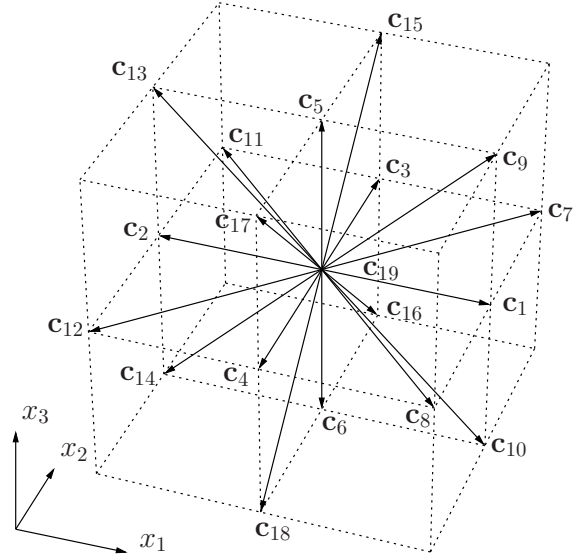}
\end{center}
\caption{D3Q19 cubic lattice with lattice vectors $\ci$.}
\label{d3q19:fig}
\end{figure}
The discretized lattice-Boltzmann (\LB) equation reads
\begin{equation}
\label{LB:disc:eq}
\ni(\x+\ci\dt,t+\dt)-\ni(\x,t) = \dt \sum_{\jj=1}^{N}S_{\ii\,\jj}\left(\nj(\x,t)-\nqj(\x,t)\right),
\end{equation}
where $\x=(\xi,\xj,\xk)$ represents a node. The discretization parameters are
$\dt$ and $\dx$, while the discrete velocities $\ci$ have the dimension
$\dx/\dt$.  The variable $\ni$ is the probable number of particles moving with
velocity $\ci$.  We use a 3D cubic lattice with 19 discrete
velocities $\ci,\,\ii=1 \ldots 19$ known
as D3Q19 (see Fig.~\ref{d3q19:fig} for a visualization)~\cite{bib:qian-dhumieres-lallemand}. The term on the
right hand side of Eq.~\eqref{LB:disc:eq} is the linearized collision
operator, where $S_{\ii\,\jj}$ is the collision
matrix also known as scattering matrix and $\nqj(\x,t)$ is the equilibrium
distribution~\cite{bib:higuera-succi-benzi}. The macroscopic density $\rho(\x,t)$ and velocity $\uvel(\x,t)$
are obtained from $\ni(\x,t)$ as
\begin{eqnarray}
\label{rho:LB:eq}
\rho(\x,t) &=& \rhor\sum_{\ii=1}^{N}\ni(\x,t),\\
\label{mom:LB:eq}
\rho(\x,t)\uvel(\x,t) &=& \rhor\sum_{\ii=1}^{N}\ni(\x,t) \, \ci,
\end{eqnarray}
where $\rhor$ is a reference density. The pressure is given by
\begin{equation}
\label{p:LB:eq}
p(\x,t) = \cs^2 \, \rho(\x,t), 
\end{equation}
with 
\begin{equation}
\label{cs:eq}
\cs = \frac{1}{\sqrt{3}}\left(\frac{\dx}{\dt}\right)
\end{equation}
being the speed of 
sound~\cite{bib:qian-dhumieres-lallemand}.
The kinematic viscosity $\nu$ is defined as 
\begin{equation}
\label{visc:LB:eq}
\nu=\frac{\cs^{2}\dt}{2}\left(2\frac{\tau}{\dt}-1\right).
\end{equation}

Within the \LB\ method, the single relaxation time (\BGK) or multirelaxation
time (\MRT) methods differ only in the way how the terms $S_{\ii\,\jj}$ and
$\nqj(\x,t)$ are defined.  In the case of the \LBBGK\
scheme~\cite{bib:bgk,bib:chen-chen-martinez-matthaeus}, the matrix
$S_{i\,j}=-\delta_{\ii\,\jj}/\tau$, where $\tau$ is a unique
relaxation time and $\delta_{\ii\,\jj}$ is the Kroneker delta.  The
equilibrium distributions
\begin{equation}
\label{qian:eq}
\nqi(\x,t)=
\wci\frac{\rho}{\rhor}
\left(1+\frac{\uveleq\cdot\ci}{\cs^{2}}+\frac{(\uveleq\cdot\ci)^{2}}
{2\cs^{4}}-\frac{\uveleq\cdot\uveleq}{2\cs^{2}} \right),
\end{equation}
are obtained
from a 2$^{\text{nd}}$ order Taylor approximation of the Maxwell
distribution function~\cite{bib:DHae:04},
with $\uveleq=\uveleq(\x,t)$ and $\rho=\rho(\x,t)$ calculated in absence of
an external force as $\uveleq(\x,t)=\uvel(\x,t)$ using
Eqs.~\eqref{rho:LB:eq} and~\eqref{mom:LB:eq}. 
The numbers $\wci$ are called lattice weights and for the D3Q19 lattice they
are
\begin{align}
\label{weight:d3q19:eq}
\wci
\begin{cases}
1/3, & \text{for } \ci=\vbf{0}, \\
1/18, & \text{for } \vert \ci \vert = 1, \\
1/36, & \text{for } \vert \ci \vert = \sqrt{2}.
\end{cases}
\end{align}
Generally, they depend on the lattice type, dimensions of space and on the
number of discrete velocities $N$. In the article of Qian
$\etal$~\cite{bib:qian-dhumieres-lallemand} a comprehensive overview on
different lattices is given.

In the case of the \LBMRT\ scheme~\cite{2002RSPTA.360..437D} 
the linearized collision operator is rewritten as 
\begin{equation}
 \sum_{\jj=1}^{N}S_{\ii\,\jj}\left(\nj(\x,t)-\nqj(\x,t)\right)
= -{\bfM}^{-1}\cdot\bfhS\cdot\left(\bfm(\x,t)-
\bfmq(\x,t)\right),
\end{equation}
i.e., it is implemented in the space of the hydraulic modes of the
problem $\bfm(\x,t)=\bfM\cdot\bfn(\x,t)$ instead of the space of
the discrete velocities, where $\bfM$ is the 
transformation matrix. 
Some modes have a hydrodynamic meaning, but some of them do
not and are used to improve the numerical stability as
described in~\cite{2000PhRvE..61.6546L,2002RSPTA.360..437D}.
The relaxation parameters of the modes 
are given in the diagonal matrix $\bfhS$. 
Only two values of the diagonal elements of $\bfhS$ 
are used to specify the
kinematic viscosity, Eq.~\eqref{visc:LB:eq}, and the bulk viscosity. These are
labeled $\tau$ and $\taub$, respectively. In summary, 
\begin{multline}
\bfhS=\mathrm{diag}(0,1/\taub,1.4,0,1.2,0,1.2,0,1.2,1/\tau,\\
1.4,1/\tau,1.4,1/\tau,1/\tau,1/\tau,1.98,1.98,1.98),
\end{multline}
is assumed for the \MRT\ method.
The other values of the diagonal elements 
are chosen to optimize the performance of the algorithm
as described in~\cite{2000PhRvE..61.6546L,2002RSPTA.360..437D}. 
The components of the equilibrium distribution
$\bfmq(\x,t)$, which represent the density and momentum are 
conserved after collision. The remaining ones are assumed to be
functions of the conserved moments and for D3Q19 are explicitly given 
in~\cite{2002RSPTA.360..437D}.

\section{Driving the fluid}
\subsection{External force in the \LB\ method}
An external force density of the fluid $\fzv(\x,t)$  is implemented by
adding a term
\begin{equation}
\varphi_{\ii}(\x,t) = \wci \frac{\dt}{{\rhor\cs}^{2}}\fzv(\x,t)\cdot\ci
\end{equation}
to the right hand side of Eq.~\eqref{LB:disc:eq}. 
The equilibrium velocity $\uveleq(\x,t)$ is then calculated using
\begin{equation}
\rho(\x,t)\uveleq(\x,t) = \rhor \sum_{\ii=1}^{N}\ni(\x,t) \, \ci,
\end{equation}
but the macroscopic velocity $\uvel(\x,t)$ has to be redefined as
\begin{equation}
\rho(\x,t)\uvel(\x,t) = \rhor \sum_{\ii=1}^{N}\ni(\x,t) \, \ci  
+ \frac{\dt}{2}\fzv(\x,t).
\end{equation}
An important point to stress here is that this implementation is just a
simplified alternative to add an external acceleration, recovering correctly
the hydrodynamic macroscopic fields with less computational
cost~\cite{PhysRevE.65.046308}. In all simulations presented in this work the
force density acts towards the direction of the $\xk$-axis, $\fzv=\fz\ek$.

\subsection{On-site Pressure Boundary Condition}
Pressure boundary conditions can be implemented as introduced by Zou and
He~\cite{bib:pf.QZoXHe.1997,HH08b}. Due to the ideal gas equation of state
setting the pressure on a specific node is equivalent to setting the density, see
Eq.~\eqref{p:LB:eq}. The on-site pressure \BC\ are used to drive the flow by
setting constant densities $\rhoin$ and $\rhoout$ ($\rhoin>\rhoout$) at the
inlet $\mc{C}(1)$ and outlet $\mc{C}(\Lxk)$ planes, i.e. the first and last
plane in the direction of the flow, respectively.  Expanding the calculation
of the density $\rho$, Eq.~\eqref{rho:LB:eq}, and momentum $\rho\uvel$,
Eq.~\eqref{mom:LB:eq}, leads to
\begin{eqnarray}
\nonumber
\frac{\rho}{\rhor} & = & \n_{1} + \n_{2} + \n_{3} + \n_{4} + \n_{5} + \n_{6}
+ \n_{7} + \n_{8} + \n_{9} + \n_{10} + \n_{11} + \\
& & \quad + \n_{12} + \n_{13} + \n_{14} + \n_{15} + \n_{16} + \n_{17} + 
\n_{18} + \n_{19}, \\
\frac{\rho}{\rhor}\ui & = & 
\n_{1} - \n_{2} + \n_{7} + \n_{8} + \n_{9} + \n_{10} - \n_{11} - \n_{12}
- \n_{13} - \n_{14}, \\
\frac{\rho}{\rhor}\uj & = & 
\n_{3} - \n_{4} + \n_{7} - \n_{8} + \n_{11} - \n_{12} + \n_{15} + \n_{16}
- \n_{17} - \n_{18}, \\
\frac{\rho}{\rhor}\uk & = & 
\n_{5} - \n_{6} + \n_{9} - \n_{10} + \n_{13} - \n_{14} + \n_{15} - \n_{16}
+ \n_{17} - \n_{18}.
\end{eqnarray}
Out of the four macroscopic values $\rho$ and $\uvel$ ($\ui$, $\uj$, and
$\uk$), only three are independent and can be set, because they have to
fulfill the mass balance equation. For the calculations performed for this
work, the components of the velocity $\ui=0$ and $\uj=0$ are being set,
resulting in a flow perpendicular to the inlet and outlet plane.  The
variables $\n_{5}$, $\n_{9}$, $\n_{13}$, $\n_{15}$, and $\n_{17}$ at the plane
$\mc{C}(1)$, and $\n_{6}$, $\n_{10}$, $\n_{14}$, $\n_{16}$, and $\n_{18}$ at
the plane $\mc{C}(\Lxk)$ (the first ones and last ones with positive and
negative component in $\xk$ direction, respectively, see
Fig.~\ref{d3q19:fig}), are not being updated by the streaming step within the
\LB\ algorithm but they have to be solved in order to obtain the requested
density (pressure) and velocity on the node.  As it is explained above, since
$\rho$, $\ui$, and $\uj$ are already set, it is not possible to also set the
velocity $\uk$, and its value has to be solved for on every node in the inlet
plane $\mc{C}(1)$ and outlet plane $\mc{C}(\Lxk)$. Therefore, we have a linear
system of four equations and six unknowns to be solved. To complete the system
of equations, bounce-back \BC\ are assumed for the non-equilibrium part of the
distribution $\ni$~\cite{bib:pf.QZoXHe.1997}
\begin{equation}
\label{rho:bc:eq}
\begin{split}
\n_{5} - \nq_{5} & = \n_{6} - \nq_{6}, \\
\n_{9} - \nq_{9} & = \n_{14} - \nq_{14} - N_{\xi}^{\xk}, \quad
\n_{13} - \nq_{13} = \n_{10} - \nq_{10} + N_{\xi}^{\xk}, \\
\n_{15} - \nq_{15} & = \n_{18} - \nq_{18} - N_{\xj}^{\xk}, \quad
\n_{17} - \nq_{17} = \n_{16} - \nq_{16} + N_{\xj}^{\xk},
\end{split}
\end{equation}
where the terms $N_{\xi}^{\xk}$ and $N_{\xj}^{\xk}$ 
are the transversal momentum corrections on the $\xk$-boundary 
for the distributions propagating in $\xi$
and $\xj$, respectively~\cite{HH08b}.
Eq.~\eqref{rho:bc:eq} then reads
\begin{equation}
\begin{split}
\n_{5} - \n_{6} & = \frac{\rho}{3\rhor}\uk, \\
\n_{9} - \n_{14} & = \frac{\rho}{6\rhor}\uk-N_{\xi}^{\xk}, \quad
\n_{13} - \n_{10} = \frac{\rho}{6\rhor}\uk+N_{\xi}^{\xk}, \\
\n_{15} - \n_{18} & = \frac{\rho}{6\rhor}\uk-N_{\xj}^{\xk}, \quad
\n_{17} - \n_{16} = \frac{\rho}{6\rhor}\uk+N_{\xj}^{\xk}.
\end{split}
\end{equation}
The on-site pressure \BC\ presented by Kutay \etal~\cite{Kutay2006381} uses
the same equations presented here, but lacks the transversal momentum
corrections $N_{\xi}^{\xk}$ and $N_{\xj}^{\xk}$. These terms are important
when velocity gradients are present~\cite{bib:pf.QZoXHe.1997}.

\section{Simulation Setup}
\begin{figure}[t]
\begin{center}
\includegraphics[width=0.6\columnwidth]{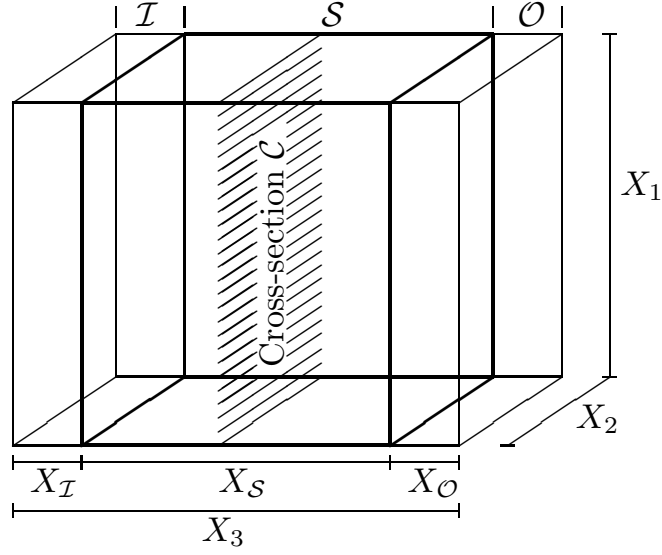}
\end{center}
\caption{Computational domain $\mc{X}$: sample $\mc{S}$ and two 
chambers $\mc{I}$ and $\mc{O}$. Cross-sections $\mc{C}$ are perpendicular
to the direction of the flow.}
\label{setup:fig}
\end{figure}

The computational domain $\mc{X}$ is composed of two subsets: $\mc{M}$
represents the solid nodes and $\mc{P}$ represents the fluid nodes, with
$\mc{M} \cup \mc{P} = \mc{X}$ and $\mc{M} \cap \mc{P}=\emptyset$. The
computational domain $\mc{X}$ with the dimensions
$[\Lxi\times\Lxj\times\Lxk]$, has three zones as presented in
Fig.~\ref{setup:fig}: the inlet $\mc{I}$, the sample $\mc{S}$, and the
outlet $\mc{O}$ with lengths $\Lxin$, $\Lxs$, and $\Lxout$, respectively.
In the case of using an injection channel the force density $\fzv$ drives
the fluid in some interval of $\xk$ within the inlet zone $\mc{I}$.
Cross-sections perpendicular to the direction of
flow are denoted by
\begin{equation}
\label{eq:Syscross}
 \mc{C}(a) =\{\x \in \mc{X} : \xk=a \}.
\end{equation}
For instance, the cross-sections $\Cinp$ and $\Coutm$ represent the first and
last cross-sections within the sample, respectively.  The average density in
the cross-section $\mc{C}(a)$ is calculated by
\begin{equation}
\label{eq:avg}
\avrg{\rho}_{\mc{C}(a)}
= \fracd{\sum_{\x \in(\mc{C}(a) \cap \mc{P})}\rho(\x)}
{\sum_{\x \in (\mc{C}(a) \cap \mc{P})}1}.
\end{equation}
The average mass flux through the whole sample is given by 
\begin{equation}
\label{massflux}
Q=\fracd{1}{\Lxk}\sum_{\x\in\mc{P}}\rho(\x)\,\uk(\x)(\dx)^{2},
\end{equation}
where $\rho(\x)\uk(\x)$ is the momentum component in direction 
of the flow.
The permeability $\kappa$ as defined by Darcy's law is
\begin{equation}
\kappa = -\nu \frac{Q}{A\,\avrg{\dpl}},
\end{equation}
where $A$ represents the cross sectional area~\cite{bib:darcy}. The average
pressure gradient in flow direction $\avrg{\dpl}$ is calculated in different
manners according to the way how the fluid is driven. This issue is explained
in more detail below.

\section{Pore-Matrix and Pore-List}
\label{matrix-list:sec}
The \LB\ method is typically implemented representing the pore space and
the solid nodes using a 3D array including the distribution functions
$\ni$ and a Boolean variable to distinguish between a pore and a matrix
node. This 3D array is known as pore-matrix. During a simulation, at every
time step the loop covering the domain includes the fluid and the solid
nodes.  Therefore, if-statements are necessary to distinguish whether the
node represents a solid node or a fluid node where the collision and
streaming steps need to be applied.  Furthermore, it has to be determined
for every node if \BC\ (periodic and no-slip) need to be applied.  The
advantage of this data structure is its straightforward implementation.
However, for the simulation of fluid flow in porous media with low
porosity the drawbacks are high memory demands and inefficient loops
through the whole simulation domain.

An alternative data structure, known as pore-list~\cite{2005ASIM:DXHHW},
uses a two-dimensional array characterizing the porous structure. It
contains the position (pore-position-list) and connectivity
(pore-neighbor-list) of the fluid nodes only. It can be generated from the
original 3D array before the first time step of the simulation, so that
only loops through the list of pore nodes not comprising any if-statements
for the lattice Boltzmann algorithm are required. The time needed to
generate and save the pore-list data is comparable to the computational
time required for one time step of the \LB\ numerical scheme implemented by
the pore-matrix. This alternative approach is slightly more complicated to
implement, but allows highly efficient simulations of flows in geometries
with a low porosity. If the porosity becomes too large, however, the
additional overhead due to the connection matrix reduces the benefits and
at some point renders the method less efficient than a standard
implementation.

In Tbl.~\ref{mem:pore:tbl} the amount of memory required for both
\LB\ implementations is presented. Introducing the notation $F$ as the
number of fluid nodes and $X^{3}=\Lxi\,\Lxj\,\Lxk$ as the total number of
nodes, we can estimate if the pore-list implementation saves memory
space using the inequality 
\begin{equation}
\label{mem:pore:eq}
  (32\bit)X^{3}+19(64\bit)X^{3}+19(64\bit)X^{2}  \geq
3(32\bit)F+18(32\bit)F+38(64\bit)F.
\end{equation}
The terms on the left hand side represent the necessary memory for the
pore-matrix implementation: pore-matrix, distribution function $\ni$, and
its buffer, respectively. The terms on the right hand side represent the
necessary memory for the pore-list implementation: pore-position-list,
pore-neighbor-list, distribution function $\ni$, and its buffer,
respectively.  Neglecting the term representing the $\ni$-buffer of the
pore-matrix implementation, because of being order $X^{2}$, we can
estimate that the pore-list implementation saves memory when $F/X^{3}
\lesssim 0.40$, where $F/X^{3}=\phi$ is the porosity of the porous medium.
\begin{table}[b]
\begin{tabular}{l l l}
  \hline 
  \multicolumn{3}{c}{Memory Space} \\
  \hline
  & \multicolumn{1}{c}{Pore-Matrix} & \multicolumn{1}{c}{Pore-List} \\
  \hline
  Geometry: & Pore-Matrix, integer $[\Lxi,\Lxj,\Lxk]$ & Pore Position-List, integer $[F,3]$ \\
  & & Pore Neighbor-List, integer $[F,18]$ \\
  $\ni$: & double-precision $[\Lxi,\Lxj,\Lxk,19]$ & double-precision $[F,19]$ \\
  $\ni$ buffer: & double-precision $[\Lxi,\Lxj,3,19]$ & double-precision
  $[F,19]$ \\
  \hline
\end{tabular}
\caption{Memory demand of implementations
\label{mem:pore:tbl}
using a pore-matrix or a pore-list data structure. $F$ represents the
number of fluid nodes.}
\end{table}
\section{Computational Domain \& Setup}
For the simulations six different lengths of the chambers $\mc{I}$ and
$\mc{O}$, $\Lxin$ and $\Lxout$ are used (see Fig.~\ref{setup:fig}). We
restrict ourselves to $\Lxin=\Lxout$ and use the lengths 20, 10, 5, 2, 1 and
0. A length of 0 represents simulations without chambers. To drive the flow
in order to generate an effective pressure gradient, we use three different
methods:
\begin{description}
\item[Injection Channel (\ICh):] In the cross-sections $\mc{C}(\Lxin-3)$,
  $\mc{C}(\Lxin-2)$, and~$\mc{C}(\Lxin-1)$ a constant force density
  $\fz\,(\dt)^2/(\rhor\,\dx) = \pwrr{-5}$ is applied and periodic \BC\ act in
  all directions. The pressure gradient is estimated using
\begin{equation}
\label{dpl:eq}
\avrg{ \dpl } = {\cs}^{2}\frac{\avrg{\rho}_{\Coutm}-\avrg{\rho}_{\Cinp}}
{(\Lxs-1)\dx}.
\end{equation}
\item[Pressure \BC\ (\pBC):] In the cross-sections $\mc{C}(1)$ and
  $\mc{C}(\Lxk)$ constant densities $\rhoin/\rhor=1+5\pwr{-5}$ and
  $\rhoout/\rhor=1-5\pwr{-5}$ are being imposed by using on-site pressure
  \BC. Periodic \BC\ act in directions $\xi$ and $\xj$.  The pressure gradient
  is calculated using Eq.~\eqref{dpl:eq}.
\item[Sample Force Density (\dpls):] A constant force density
  $\fz\,(\dt)^2/(\rhor\,\dx)=\pwrr{-5}$ acts inside the sample $\mc{S}$. In
  this case it is not necessary to estimate the average pressure gradient
  because its value is simply given by $\fz$.
\end{description}
\begin{figure}[t]
\hfill
\begin{minipage}[t]{0.3\textwidth}
\begin{center}
Injection Channel 

(\ICh)

\includegraphics[width=1.0\columnwidth]{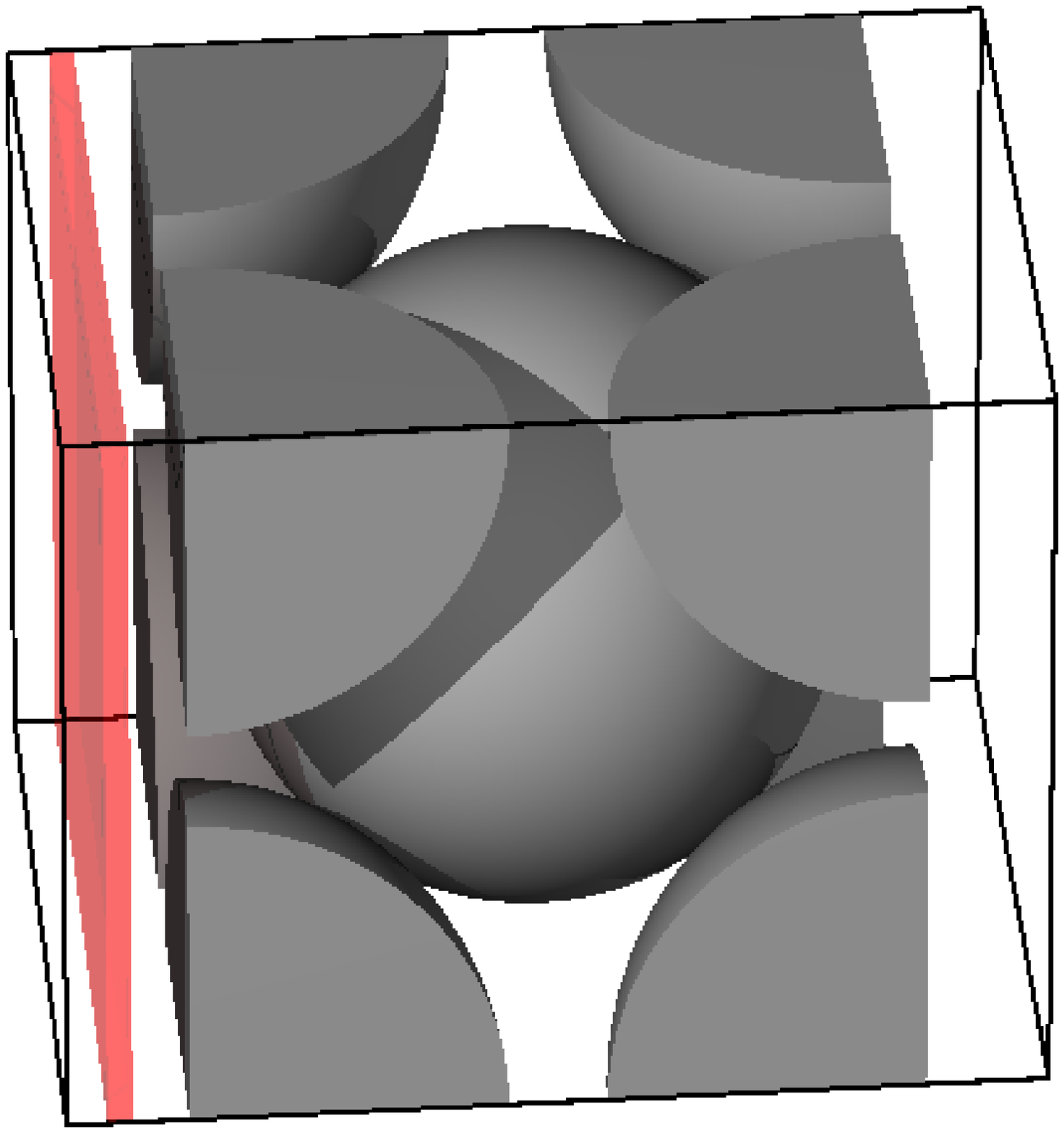}
\end{center}
\end{minipage}
\hfill
\begin{minipage}[t]{0.3\textwidth}
\begin{center}
Pressure \BC\ 

(\pBC)

\includegraphics[width=1.0\columnwidth]{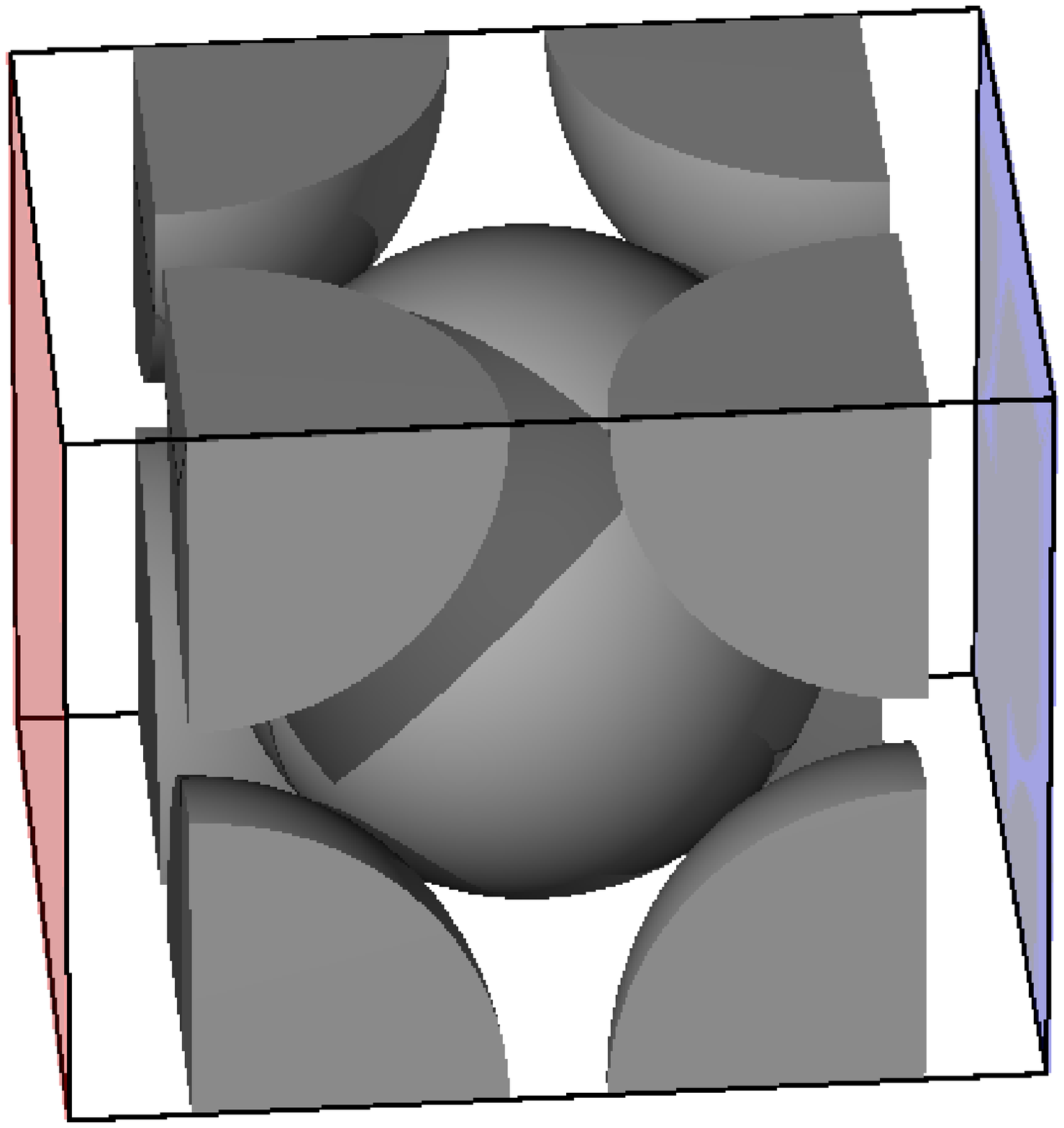}
\end{center}
\end{minipage}
\hfill
\begin{minipage}[t]{0.3\textwidth}
\begin{center}
Sample Force Density 

(\dpls)

\includegraphics[width=1.0\columnwidth]{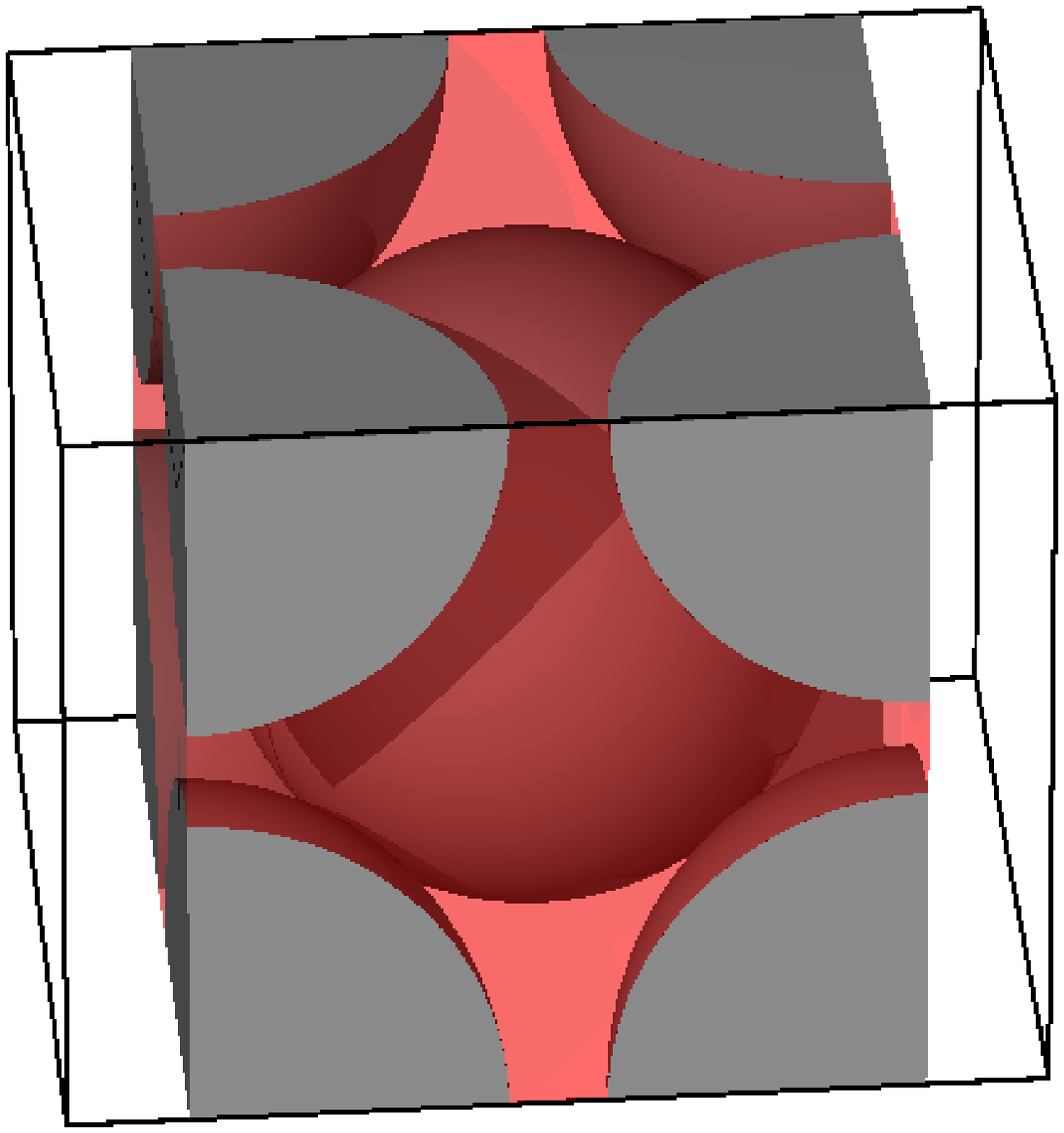}
\end{center}
\end{minipage}
\hfill
\caption{Visualization of the different methods to drive the fluid.
  The sample is represented as a \BCC\ sphere array.
  \ICh: the marked zone within $\mc{I}$ is where the fluid is driven by the
  force density. \pBC: at the marked cross-sections $\mc{C}(1)$
  and $\mc{C}(\Lxk)$ constant densities are being set. \dpls: inside the
  sample (marked zone) the fluid is driven by the applied force.}
\label{comp:domain:fig}
\end{figure}

Two different samples are being used, the first one is a 3D Poiseuille flow in
a square pipe of side length $\dd$. The sample dimensions are
$\Lxj=\Lxk=\dd/\dx+2$, and $\Lxs=50+\Lxin+\Lxout$, where the cross-sectional
area in this case is $A=\dd^{2}$. The two extra nodes of $\Lxj$ and $\Lxk$
are used to provide solid walls which define the pipe. The theoretical
permeability for 3D Poiseuille flow in a square pipe is given
by~\cite{bib:jcis.TPaDSi.2000}
\begin{equation}
\kappa^{\theo} = \frac{\dd^{2}}{4}
\left(\frac{1}{3}-\frac{64}{\pi^{5}}
\sum_{n=0}^{\infty}\frac{\tanh\left((2n+1)\fracd{\pi}{2}\right)}
{(2n+1)^{5}}\right) \approx 0.03514\,\dd^{2}.
\end{equation}
The second sample is composed of a cube filled with a \BCC\ sphere array. The
sample geometry is adjusted by varying the radius of the spheres. The
dimensions are $\Lxj=\Lxk=50$ and $\Lxs=50$ with $\Lxi=\Lxs+\Lxin+\Lxout$,
giving a cross-sectional area of $A=2500\,(\dx)^{2}$. The theoretical value
for the permeability is given by~\cite{Hasimoto1959,Sangani1982343}
\begin{equation}
\kappa^{\theo} = \frac{a^{2}}{6 \pi\, \chi\, h},
\end{equation}
where $a$ and $h$ represent the side length of the cube and the inverse of the
dimensionless drag, respectively. As it is presented
in~\cite{bib:cf.CPaLLuCMi.2006}, $h$ is entirely determined by the geometry
characteristic of the sphere array and can be represented as a series
expansion of the radius ratio
\begin{equation}
\chi=\frac{r}{r_{\max}},
\end{equation}
where $r$ represents the radius of the spheres and $r_{\max}=\sqrt{3}\,a/4$
the maximum radius that the spheres can have in the \BCC\ array until
they touch each other.
The relative error calculated by
\begin{equation}
\epsilon(\kappa)=\frac{\vert \kappa-\kappa^{\theo} \vert}{\kappa^{\theo}},
\end{equation}
is used to qualify the results of our simulations.

\section{Results}
In order to compare the efficiency and accuracy of different implementations
to compute permeabilities, its value is being estimated by applying the
alternatives to drive the flow presented above, i.e., Injection Channel
(\ICh), pressure \BC\ (\pBC), and Sample Force Density (\dpls).
It is well know that a slight dependency of the permeability in dependence on
the relaxation time $\tau$ and thus the fluid viscosity can be observed if
standard bounce back boundary conditions are used. Using \LBMRT\ method the
$\kappa$-$\tau$ correlation is significantly smaller than for
\LBBGK~\cite{bib:ginzburg-dhumieres,2002RSPTA.360..437D}. To keep focus in
the accuracy comparison of the different setups, we use the same relaxation
times for all presented calculations. The values are set to $\tau/\dt=0.857$
and $\taub/\dt=0.84$, because we found them to produce very accurate results
for 3D Poiseuille flow in a previous contribution~\cite{nzrhh2010:bib}.
The first test
geometry for measuring the permeability is a 3D Poiseuille flow with different
pipe lengths $d/\dx$.
\begin{figure}[t]
\begin{minipage}[t]{0.475\textwidth}
\centering Pressure \BC\ (\pBC)

\includegraphics[width=1.0\textwidth]{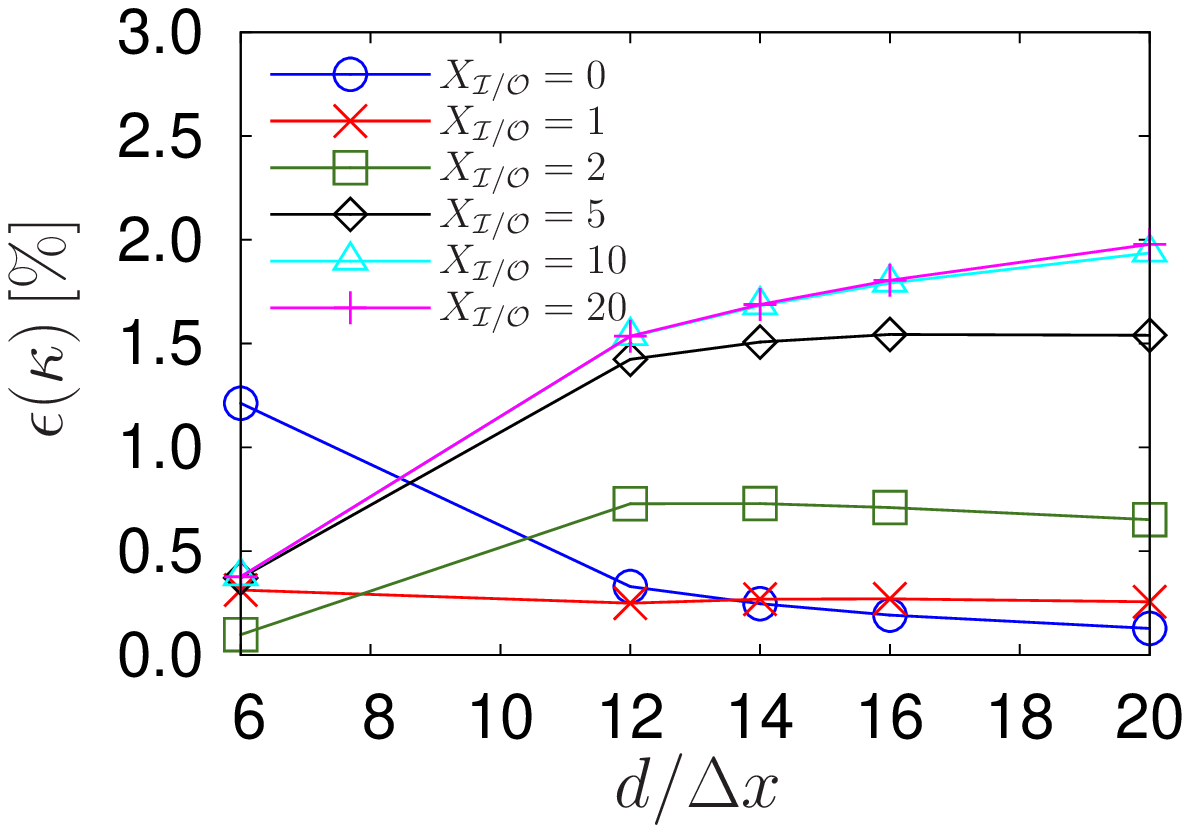}
\end{minipage}
\hfill
\begin{minipage}[t]{0.475\textwidth}
\centering Sample Force Density (\dpls)

\includegraphics[width=1.0\textwidth]{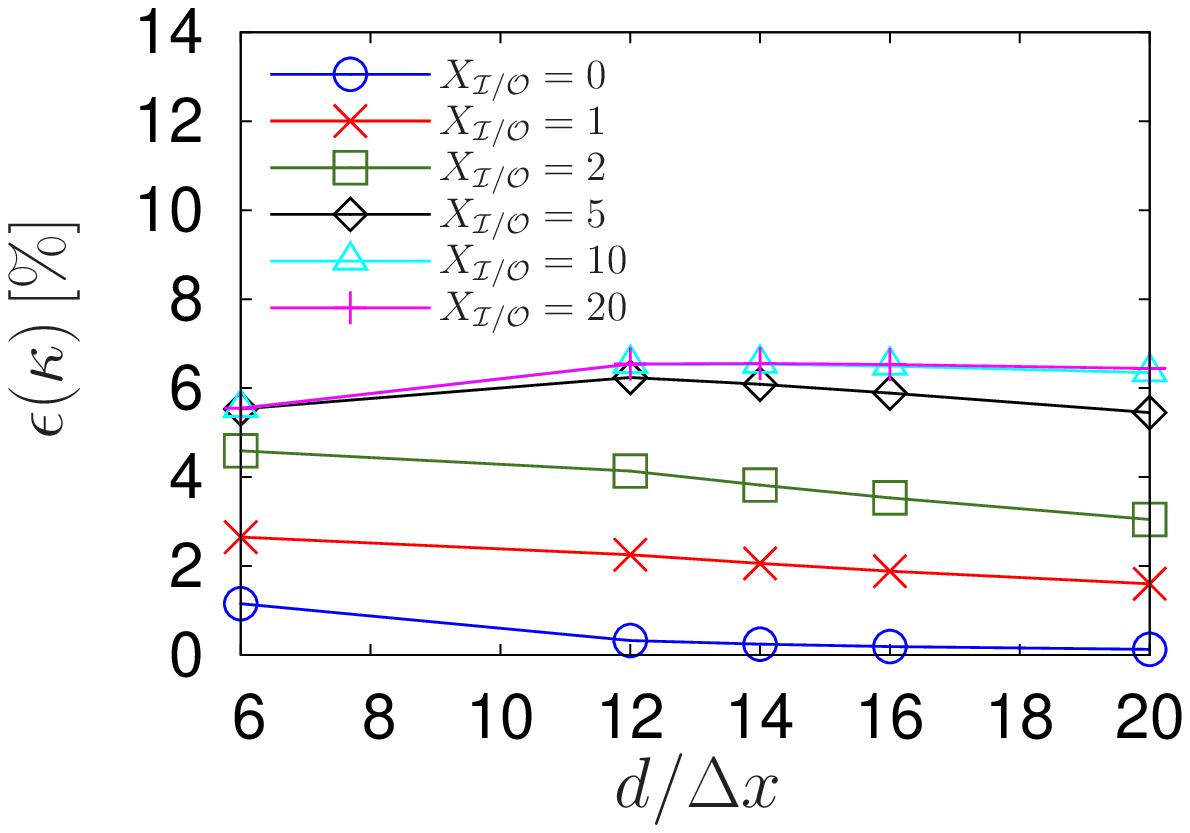}
\end{minipage}
\hfill

\hfill
\begin{minipage}[t]{0.475\textwidth}
\centering Injection Channel (\ICh)

\includegraphics[width=1.0\textwidth]{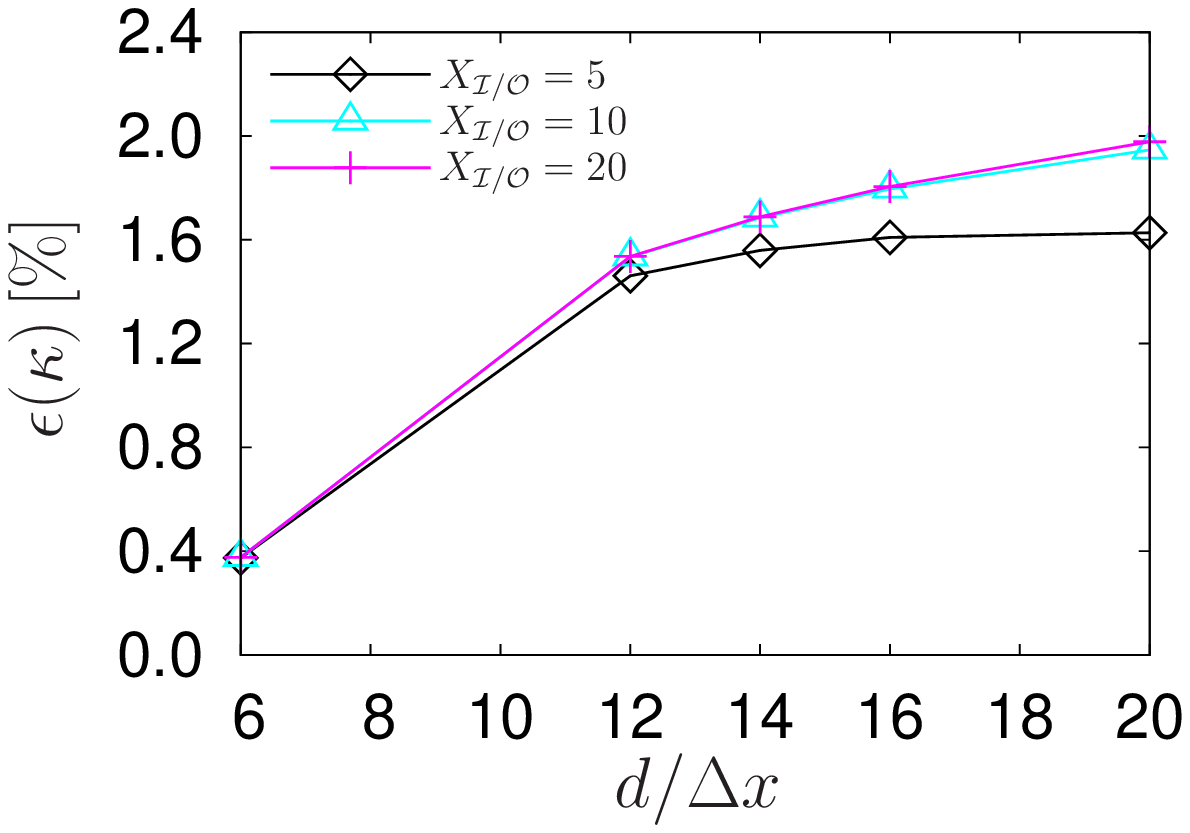}
\end{minipage}
\hfill
\begin{minipage}[t]{0.475\textwidth}
\caption{Relative error of permeability of a 3D Poiseuille flow in a
  square pipe. Three different alternatives to drive the flow \pBC, \dpls\
and \ICh\ are tested together with different lengths of the chambers
$\mc{I}$ and $\mc{O}$. Using the very popular alternative \dpls, the
results are highly dependent on the length of the chambers. Therefore,
\dpls\ is not taken into account for any further studies.} 
\label{poiseuille:3D:fig}
\end{minipage}
\hfill
\end{figure}
Fig.~\ref{poiseuille:3D:fig} shows the relative error of the computed
permeabilities for chambers of different length and the three different setups
presented. In case of \pBC\ and \ICh\ the results are very accurate and the
error stays below 2.5\%. For \dpls, however, there is a strong dependency of
the results on the length of the chamber. If the sample is not periodic as it
would be for experimentally relevant samples such as a discretization of a
sandstone, it is not possible to avoid the chambers completely. In fact, the
chambers are then absolutely needed providing a fluid reservoir before and
after the sample, but as can be seen in Fig.~\ref{poiseuille:3D:fig}~\dpls,
they produce a massive loss in accuracy. For this reason we do not take into
account the \dpls\ setup for further calculations. Our finding is of
particular interest because this setup is the most popular one in the
literature on permeability measurements using the \LB\ method.

On the other hand, for the results obtained by alternatives \pBC\ and
\ICh, one should take into account that the major error in permeability
estimation of a real sample with the \LB\ method is due to the
discretization of the porous space. For the 3D Poiseuille flow in a square
pipe this error is minimal since the pipe is aligned with the lattice.
Therefore, a systematic error of the order of 2\% is satisfactory.

\begin{figure}[t]
\hfill
\begin{minipage}[t]{0.475\textwidth}
\centering $\bepsilon(\x,0,2)$

\includegraphics[width=1.0\textwidth]{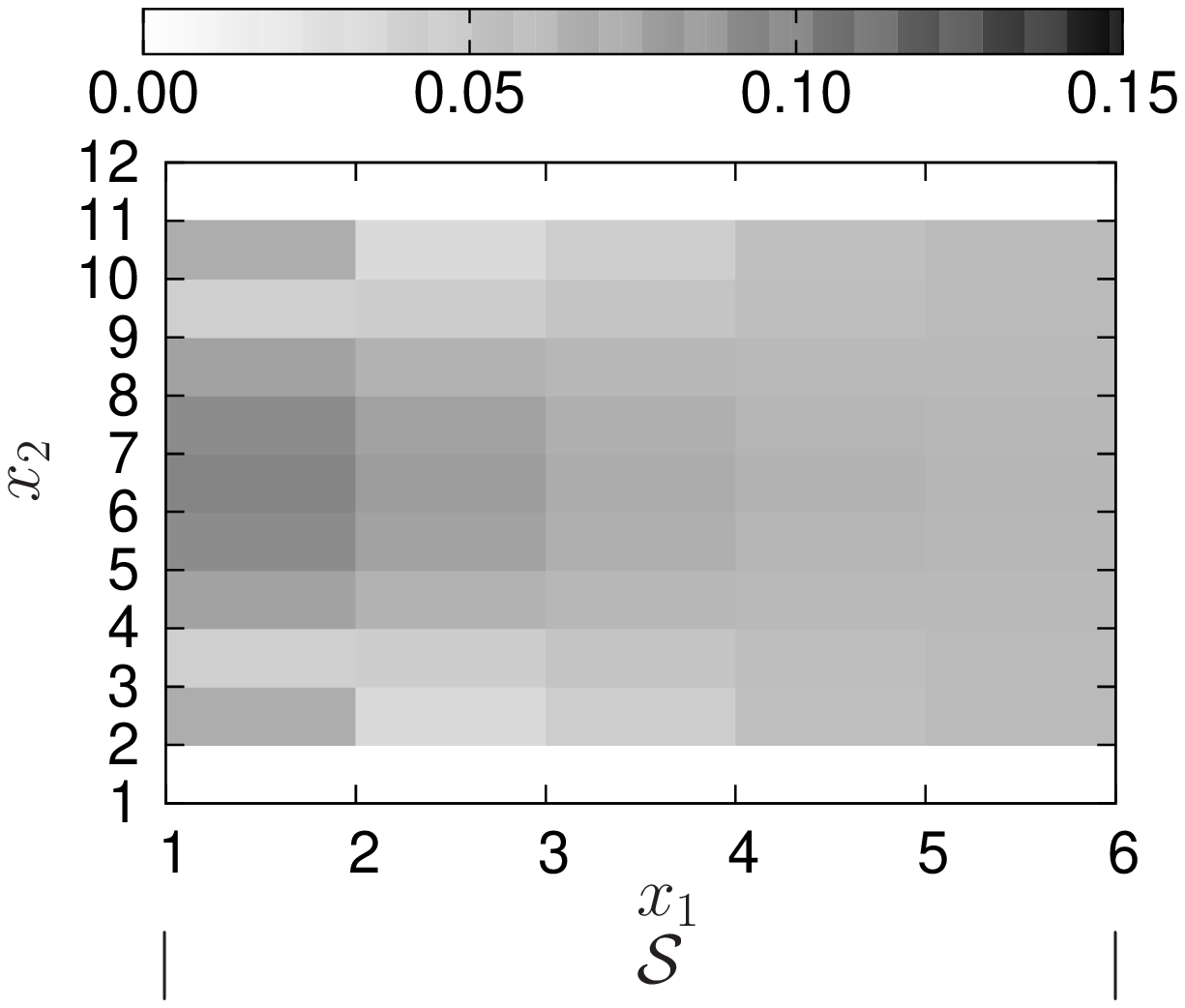}
\end{minipage}
\hfill
\begin{minipage}[t]{0.475\textwidth}
\centering $\bepsilon(\x,2,10)$

\includegraphics[width=1.0\textwidth]{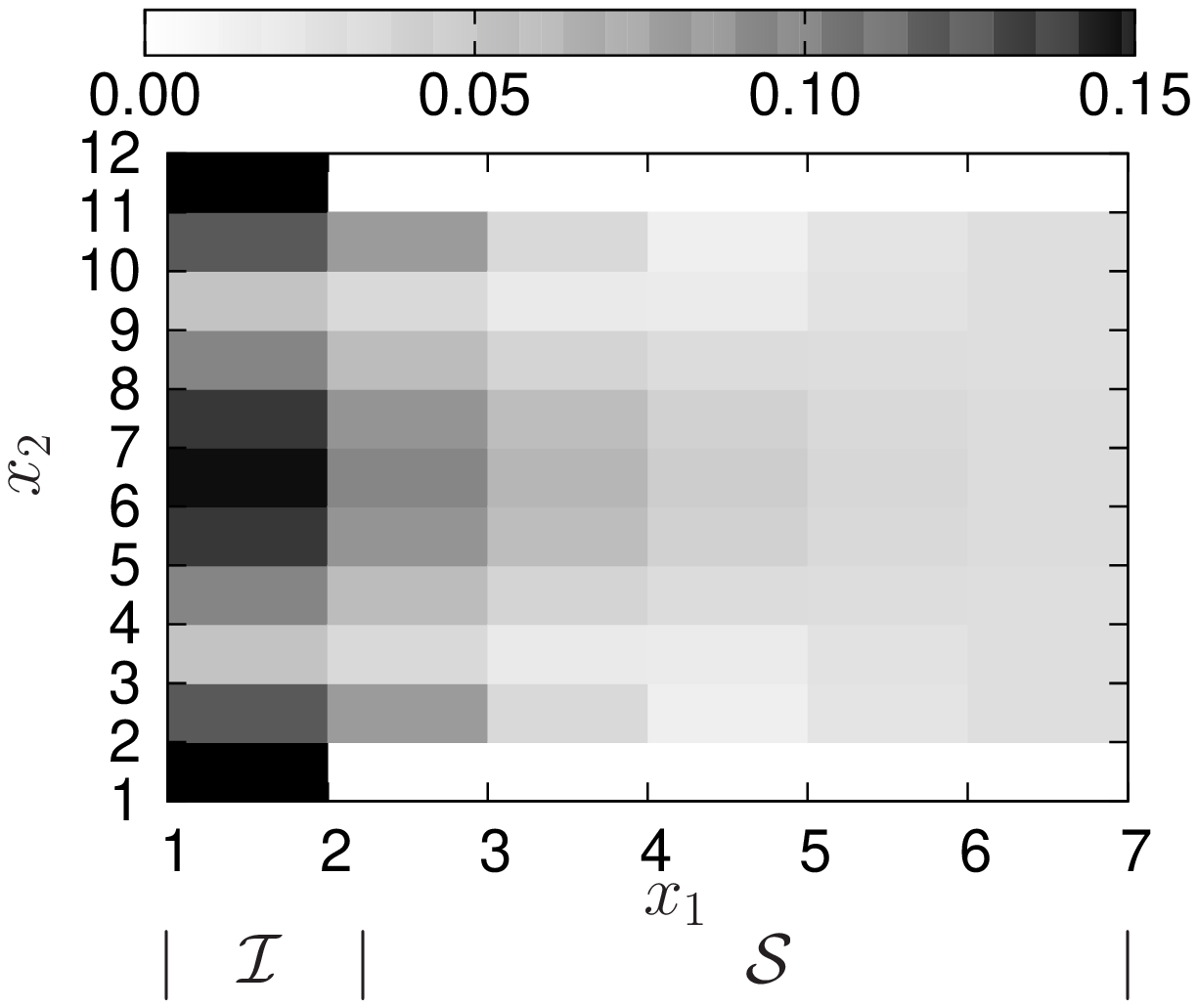}
\end{minipage}
\hfill
\caption{Two examples comparing the difference in the velocity field $\ui$
  obtained in a 2D Poiseuille flow $d/\dx=10$ 
  when the fluid is driven using \dpls\ and two different chamber
  lengths $\Lxinout$ are used. The difference is calculated using the relative
  error field $\bepsilon(\x,A,B)$ presented in Eq.~\eqref{bepsilon:eq}. For
  both examples, the velocity field $\ui$ is only comparable in the zone
  that both domains have in common and which is defined by the shortest chamber.
  On the $\xj$-axis the first and last nodes inside the sample
  $\mc{S}$ represent the solid walls defining the pipe. Although the major
  difference resides at the inlet, a constant relative error remains
inside the sample, which is responsible for the difference in the measured permeability.}
\label{ux:dpls:fig}
\end{figure}

To estimate the effect of the chamber length on the permeability results when
the \dpls\ setup is being used, we simulate a 2D Poiseuille flow to compare
the velocity field $\ui$ obtained inside the sample when chambers of two
different lengths are used. To visualize the difference a relative error field
$\bepsilon$ is calculated as
\begin{equation}
\label{bepsilon:eq}
  \bepsilon(\x,A,B) = 
\frac{\vert \ui(\x,\Lxinout=A) - \ui(\x,\Lxinout=B) \vert}
{\ui(\x,\Lxinout=A)},
\end{equation}
where $A$ and $B$ represents the two chamber lengths whose velocity field
$\ui$ are being compared. 
Fig.~\ref{ux:dpls:fig} displays two examples comparing the results of
$\Lxinout=0$ and $\Lxinout=2$, and the results of $\Lxinout=2$ and
$\Lxinout=10$.
As can be clearly seen, the difference of permeability estimation when using
different chamber lengths, can be explained
by the influence of the chamber geometry on the velocity fields $\ui$. 
Not only the velocity field entering the sample is dependent on the chamber
length, but also the velocity field inside the sample, where the parabolic
profile is already adopted. This issue is clearly shown in both examples of
Fig.~\ref{ux:dpls:fig}, where a constant relative difference in the velocity
field $\ui$ remains up to the fifth node inside of the sample.
It is actually this difference, which responds to how
the fluid enters the sample, that is responsible for the dependency of the
permeability on the length of the chamber. This clearly explains the
results shown in
Fig.~\ref{poiseuille:3D:fig}~\dpls.

\begin{figure}[t]
\centering  Pressure \BC\ (\pBC)

\hfill
\begin{minipage}[t]{0.475\textwidth}
\centering \LBBGK\

\includegraphics[width=1.0\textwidth]{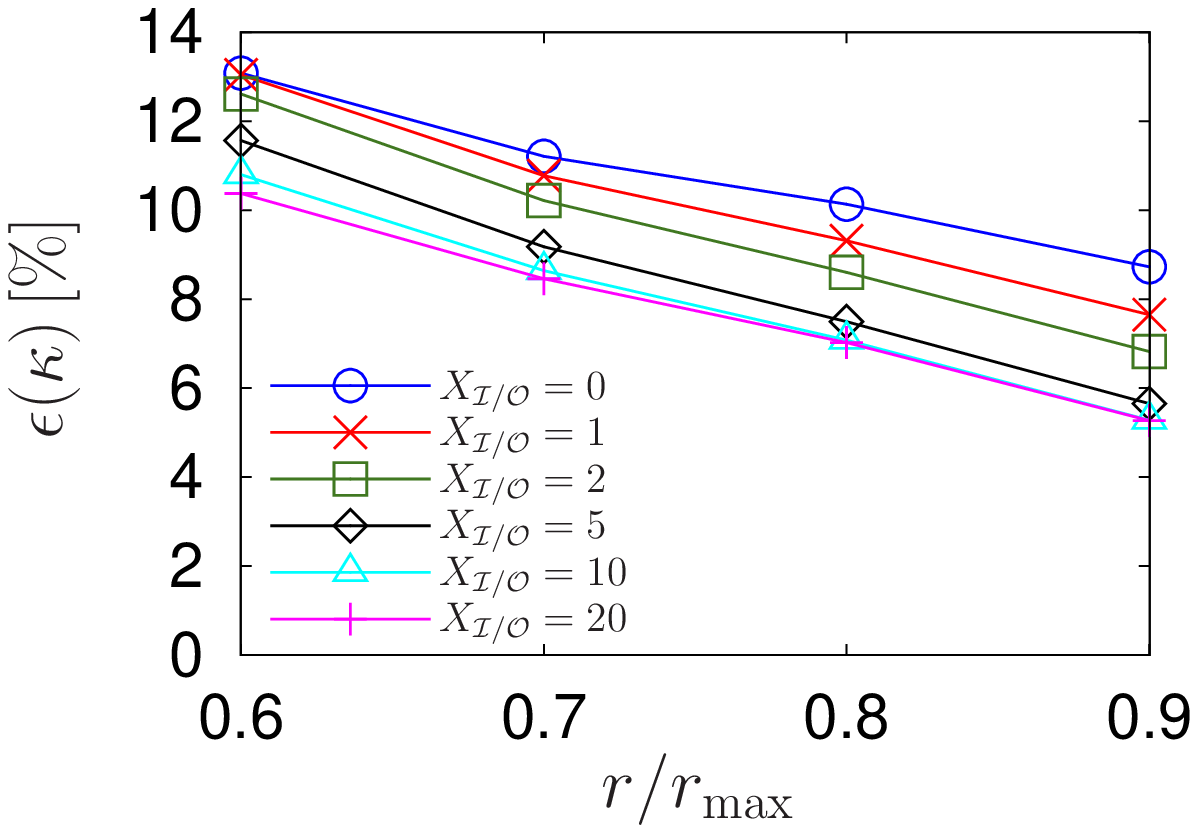}
\end{minipage}
\hfill
\begin{minipage}[t]{0.475\textwidth}
\centering \LBMRT\

\includegraphics[width=1.0\textwidth]{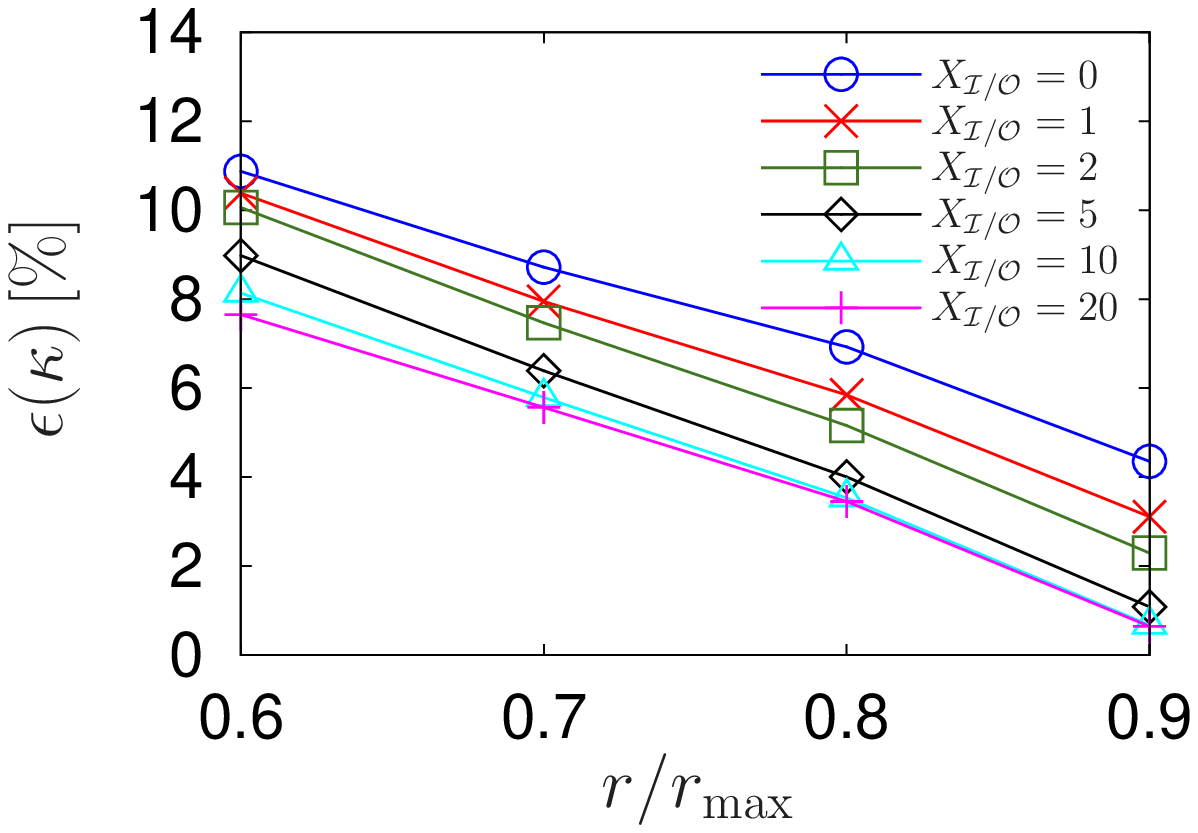}
\end{minipage}
\hfill

\centering Injection Channel (\ICh)

\hfill
\begin{minipage}[t]{0.475\textwidth}
\centering \LBBGK\

\includegraphics[width=1.0\textwidth]{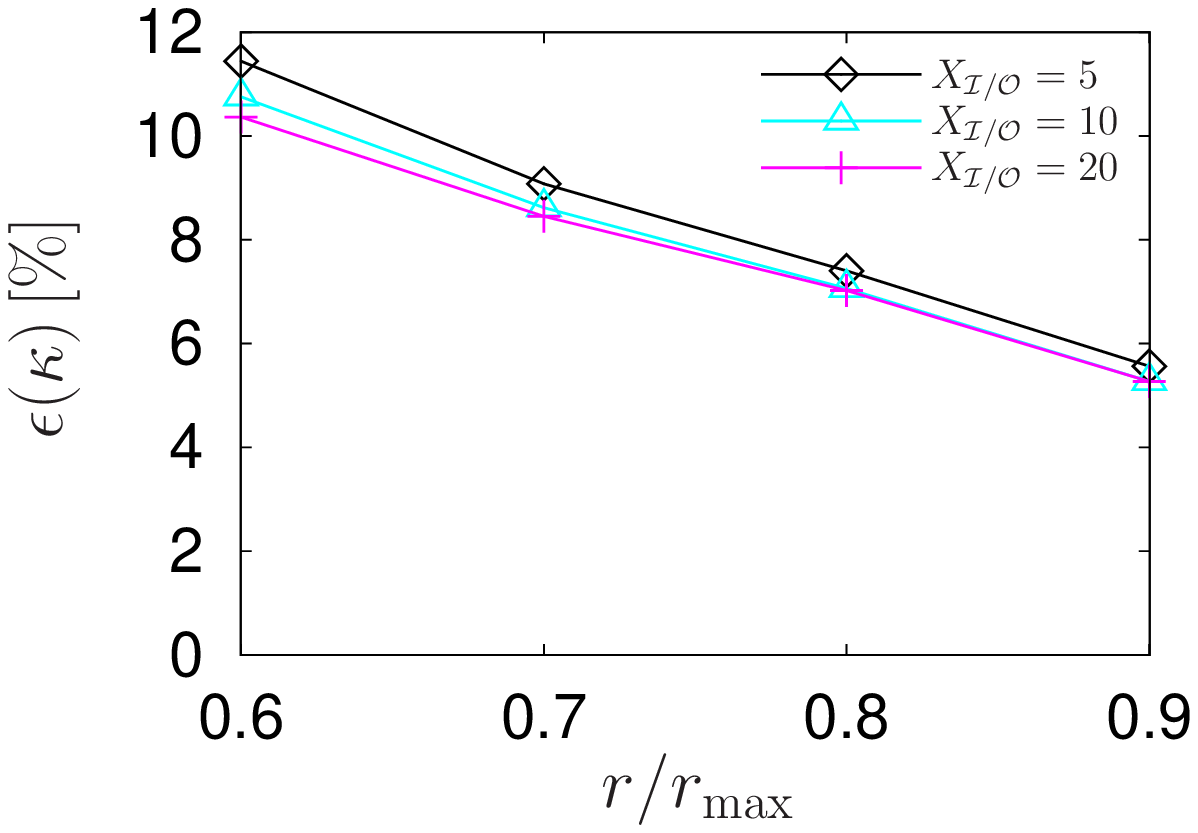}
\end{minipage}
\hfill
\begin{minipage}[t]{0.475\textwidth}
\centering \LBMRT\

\includegraphics[width=1.0\textwidth]{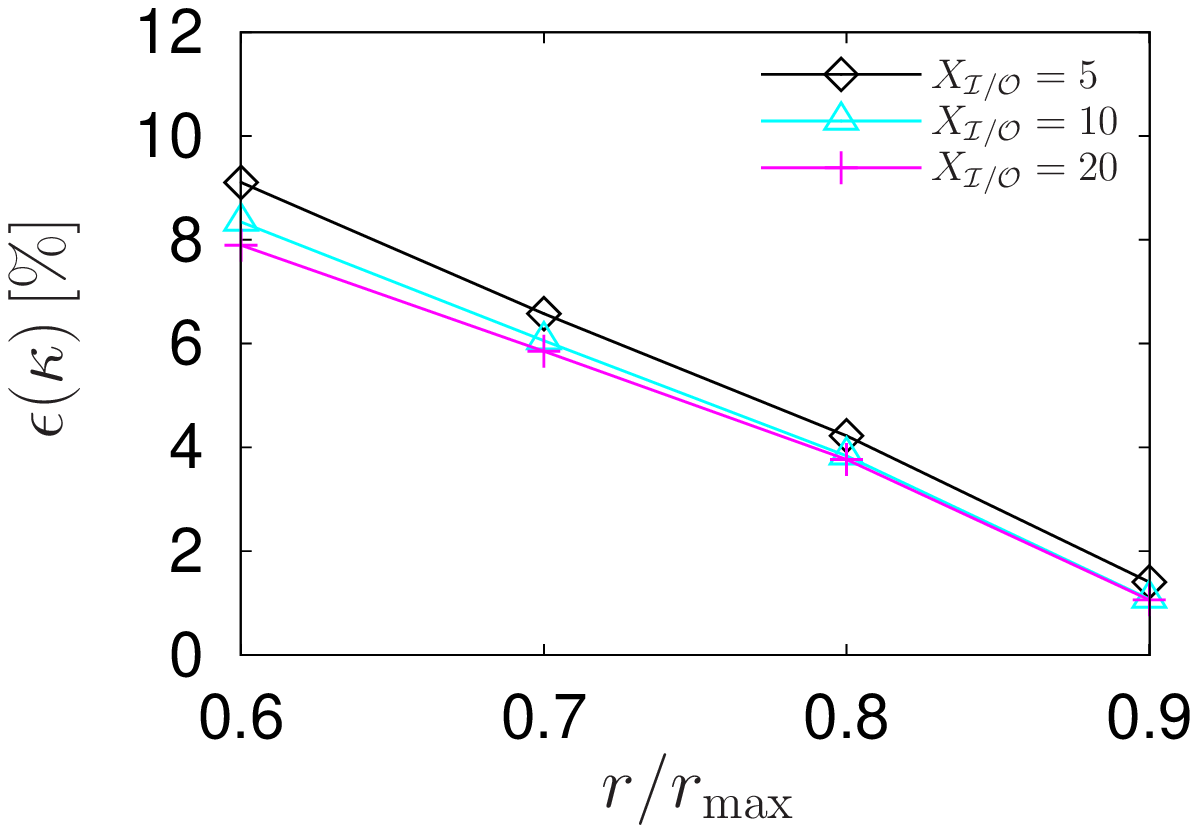}
\end{minipage}
\hfill
\caption{Relative error of permeability for a \BCC\ cubic array of
spheres.  The alternatives \pBC, \ICh\ are being tested together with
different lengths of the chambers $\mc{I}$ and $\mc{O}$ and also using the
\LBBGK\ and \LBMRT\ method. The \LBMRT\ together with \pBC\ without
chambers achieves the same accuracy as using \LBBGK\ with \ICh, but in the
latter case chambers are required.}
\label{k:bcc:fig}
\end{figure}

Fig.~\ref{k:bcc:fig} displays the permeability calculated for a \BCC\
cubic array using the \LBBGK\ and \LBMRT\ method and the alternatives
\pBC\ and \ICh\ with different chamber length. In contrast to the 3D
Poiseuille flow, the measurements suffer indeed from an error
due the domain discretization. As in Fig.~\ref{poiseuille:3D:fig} for the
3D Poiseuille flow, the dependency of the results using \ICh\ on the
length of the chambers is narrower than using \pBC\ for both \LB\ methods.

As it was stressed above, typically the major contribution to the error of
permeabilities estimated using the \LB\ method is the discretization. This
issue is clearly seen in Fig.~\ref{k:bcc:fig} where the relative error in
the permeability estimation increases if the radius of the sphere is
decreased, which is equivalent to reducing the resolution.

The \LBMRT\ method in general produces more accurate results than \LBBGK.
The best results can be obtained when using \LBMRT\ together with \ICh\
and $\Lxinout=20$. However, reducing the channel length does not have a
strong influence. On the other hand, if we compare the results yield by
using \ICh\ with the ones obtained with \pBC, only minor differences can
be obtained for identical chamber lengths.
Following this we can assert that
there is no significant difference in accuracy when using either \ICh\ or
\pBC, as it was also expected from the data presented in
Fig.~\ref{poiseuille:3D:fig}. The parameter impacting on the accuracy is
the chamber length. The \LBMRT\ together with \pBC\ without chambers
achieves the same accuracy as \LBBGK\ with \ICh, but in the latter case
chambers are required.  Since long chambers increase the computational
domain, they are a burden in terms of computational effort. Thus, it is
important to understand their influence in order to decide whether a lower
accuracy can be accepted if the computational effort is substantially
reduced.

The efficiency of the \LB-implementations, i.e., pore-matrix and pore-list
as their memory demands explained above are directly related to the porosity
of the sample. 
It is important to stress that the efficiency is independent of the setup used
for the permeability estimation but it depends on the \LB\ method used, i.e.,
\BGK\ or \MRT. In the case of using \LBBGK\ instead of \LBMRT\ the CPU time
required decreases by 15\%--20\%~\cite{bib:cf.CPaLLuCMi.2006}.

The use of chambers and more precisely their lengths determine the porosity of
the whole computational domain. For example, if in a cubic sample representing
a realistic porous medium with porosity $\phi=10\%$ we add chambers with
lengths 5\% of the sample length, we obtain a computational domain with
porosity $\phi\approx 18.2\%$. 
Even more dramatically is the case if the sample
has a smaller porosity. If we add chambers with the same length explained
above to a cubic sample with porosity $\phi=5\%$, the total porosity would
increase by a factor of $\sim\!\!\!173\%$. 
Since the efficiency of both \LB-implementations scales linearly with the
porosity, the use of chambers is expensive in terms of computational cost and
should be avoided if possible.
For low porosity values, as realistic samples have, the pore-list
implementation is more efficient than pore-matrix. 
On the other hand, for more open domains, as pipes for example,
the pore-matrix data structure is more efficient since it does not require
the connection matrix.
A quantitative comparison of the computation efficiency is not easily
possible. Even though it is possible to calculate for both implementations
the number of floating-point operations necessary to perform a single time
step, a quantitative analysis of the performance of two idependent codes
is not as straightforward: the code performance strongly depends on the
actual implementation details, memory access patterns and in particular
also on the cache performance of the used architecture. 

\section{Conclusions}
We presented different alternative setups to estimate the permeability of a
porous medium, which utilize an injection channel (\ICh), pressure
boundary conditions (\pBC), or a force density applied in the sample
volume (\dpls). The setups differ on how the fluid is driven and how the
pressure gradient is being estimated. Further, the accuracy of the \LBBGK\
and \LBMRT\ method together with different setups with variable chamber
length, has been investigated. Another point that has been stressed is
the efficiency of the \LB-implementations comparing the standard
implementation using a pore-matrix data structure with a pore-list code to
represent the geometry.

Our findings show that the use of \dpls\ as an alternative for permeability
estimation should not be taken into account due to the need of chambers as
fluid reservoir together with the strong dependency of the results on the
length of these chambers. This is an important result since driving the fluid
by using a body force applied in the sample volume is probably the most
popular approach in the literature on permeability calculations using the \LB\
method.

Further, the pore-list implementation allows to reduce the computational
effort required to simulate fluid flows in porous media substantially if
the porosity of the porous medium is small.

Finally, taking into account the extra computational effort added because
of the number of required lattice nodes when chambers are used, it is highly
recommended to completely eliminate them by using \pBC.  A resulting
compromise in the accuracy can be resolved by using the \LBMRT\ method.
For this reason in the case of facing the permeability estimation of a
realistic media, pore-list, \pBC, and \LBMRT\ is the combination suggested
by the authors taking into account the computational domain size, accuracy
and required computer time.

\section*{Acknowledgments}
We would like to thank the ``Deutscher Akademischer Austausch Dienst
(DAAD)''
and the ``Stichting voor de Technische Wetenschappen (STW)/ Nederlandse
Organisatie voor Wetenschappelijk Onderzoek (NWO)'' for financial support.
Martin Hecht, Rudolf Hilfer, Thomas Zauner, and Frank Raischel are highly
acknowledged for fruitful discussions.


\end{document}